\documentclass[aip,reprint,twocolumn,amssymb,amsmath,superscriptaddress]{revtex4-1}
\usepackage{graphicx,color,colortbl}
\usepackage{epstopdf}
\usepackage{bm}
\usepackage{eucal}
\usepackage{mathrsfs}
\newcommand{\red}{\textcolor{red}}

\begin{document}

\title{Cross frequency coupling in next generation inhibitory neural mass models}

\author{Andrea Ceni}
\affiliation{Department of Computer Science, College of Engineering, Mathematics and Physical Sciences, University of Exeter, Exeter EX4 4QF, UK}
\author{Simona Olmi}
\affiliation{Inria Sophia Antipolis M\'{e}diterran\'{e}e Research Centre, 2004 Route des Lucioles, 06902 Valbonne, France}
\affiliation{CNR - Consiglio Nazionale delle Ricerche - Istituto dei Sistemi Complessi, via Madonna del Piano 10, 50019 Sesto Fiorentino, Italy}
\author{Alessandro Torcini}
\affiliation{Laboratoire de Physique Th\'eorique et Mod\'elisation, Universit\'e de Cergy-Pontoise, CNRS, UMR 8089,
95302 Cergy-Pontoise cedex, France}
\affiliation{CNR - Consiglio Nazionale delle Ricerche - Istituto dei Sistemi Complessi, via Madonna del Piano 10, 50019 Sesto Fiorentino, Italy}
\author{David Angulo-Garcia}
\affiliation{Grupo de Modelado Computacional - Din\'{a}mica y Complejidad de Sistemas. Instituto de Matem\'{a}ticas Aplicadas. 
Carrera 6 \#36 - 100. Universidad de Cartagena. Cartagena de Indias, Colombia}

\begin{abstract}
Coupling among neural rhythms is one of the most important mechanisms at
the basis of cognitive processes in the brain. In this study we consider a neural mass model, rigorously obtained from the microscopic dynamics of an inhibitory spiking network with exponential synapses, able to autonomously generate collective oscillations (COs). These oscillations emerge via a super-critical Hopf bifurcation, and their frequencies are controlled by the synaptic time scale, the synaptic coupling and the excitability of the neural population. Furthermore, we show that two inhibitory populations in a master-slave configuration with different synaptic time scales can display various collective dynamical regimes: namely, damped oscillations towards a stable focus, 
periodic and quasi-periodic oscillations, and chaos. 
Finally, when bidirectionally coupled the two inhibitory populations 
can exhibit different types of $\theta$-$\gamma$ cross-frequency couplings (CFCs): 
namely, phase-phase and phase-amplitude CFC.
The coupling between $\theta$ and $\gamma$ COs is enhanced in presence of a 
external $\theta$ forcing, reminiscent of the type of modulation induced
in Hippocampal and Cortex circuits via optogenetic drive.
\end{abstract}

\maketitle

\begin{quotation}
In healthy conditions, the brain's activity reveals 
a series of intermingled oscillations, generated by large ensembles of neurons, 
which provide a functional substrate for information processing. How single neuron properties influence neuronal population dynamics is an unsolved question, whose solution could help in the understanding of the emergent collective behaviors 
arising during cognitive processes. Here we consider a neural mass model, which reproduces exactly the macroscopic activity of 
a network of spiking neurons. This mean-field model is employed to shade some light on an important and ubiquitous neural mechanism underlying information processing in the brain: the $\theta$-$\gamma$ cross-frequency coupling.  In particular, we will explore in detail the conditions under which two coupled inhibitory neural populations can generate these functionally relevant coupled rhythms.
\end{quotation}

\section{Introduction}

Neural rhythms are the backbone of information coding in the brain 
\cite{Buzsaki2006Rhythms}. These rhythms are the proxy of large populations 
of neurons that orchestrate their activity in a regular fashion and selectively 
communicate with other populations producing complex 
functional interactions \cite{varela2001brainweb,canolty2010functional}.

One of the most prominent rhythmic interaction appearing in the brain is the so-called $\theta$-$\gamma$ coupling, exhibited between the slow oscillating $\theta$ band (4Hz - 12Hz) and the faster $\gamma$ rhythm (25 -100 Hz) \cite{lisman2013theta}. This specific frequency interaction is an example
of a more general mechanism termed Cross-Frequency-Coupling (CFC)
\cite{jensen2007cross,canolty2010functional}.
CFC has been proposed to be at the basis of sequence representation, long distance communication, sensory parsing and de-multiplexing \cite{jensen2007cross,hyafil2015neural}. CFC can manifest in a variety of ways depending on the type of modulation that one rhythm imposes on the other (Amplitude-Amplitude, Amplitude-Frequency, Phase-Amplitude, Frequency-Frequency, Frequency-Phase or Phase-Phase) \cite{jensen2007cross}. 

In this article, we will focus on Phase-Phase (P-P) and Phase-Amplitude (P-A) couplings of $\theta$ and $\gamma$ rhythms.
In particular, P-P coupling  refers to n:m phase locking between gamma and theta phase oscillations \cite{rosenblum2000} and it has been demonstrated to play a role in visual tasks in humans \cite{holz2010theta}  and it has been identified 
in the rodent hippocampus during maze exploration \cite{belluscio2012cross}.
The P-A coupling (or $\theta$-nested $\gamma$ oscillations) corresponds to the fact that the phase of the theta-oscillation modifies the amplitude of the gamma waves and it has been shown to support the formation of new episodic memories in the human hippocampus \cite{lega2014} and to emerge in various part of the rodent brain during optogenetic theta stimulations {\it in vitro} \cite{akam2012,pastoll2013,butler2016,butler2018}.

In an attempt to understand the complex interactions emerging during CFC, several mathematical models describing  the activity of a large ensemble of neurons have been proposed \cite{david2003neural}. 
Widely known examples include the phenomenologically based Wilson-Cowan \cite{wilson1972excitatory} and Jansen-Rit \cite{jansen1995electroencephalogram} neural mass models. While the 
usefulness of these heuristic models is out of any doubt, they are not related to
any underlying microscopic dynamics of a neuronal population. A second approach consists on the use of a mean-field (MF) description of the activity of a
large ensemble of interacting neurons \cite{Ricciardi1971, treves1993, BrunelHakim1999, Brunel2000Sparse, richardson2010firing, moreno2010response, olmi2017exact}. Although in this latter MF framework 
some knowledge is gained on the underlying microscopic dynamics that manifests at the macroscopic level, it comes with the downside of several, not always verified, assumptions about the statistics of single neuron firings \cite{ekke2019}.

Recently, it has been possible to obtain  in an
exact manner a macroscopic description of an infinite ensemble of 
pulse-coupled Quadratic Integrate-and-Fire (QIF) neurons \cite{luke2013,pazo2014,montbrio2015macroscopic,coombes2019}. 
This result has been achieved thanks to the Ott-Antonsen ansatz, which allows to exactly obtain the macroscopic evolution 
of phase oscillator networks fully coupled via purely sinusoidal field \cite{ott2008low}. In particular, in \cite{montbrio2015macroscopic} the authors have been able to derive for QIF neurons with instantaneous synapses an exact neural mass model in terms of the firing rate and the average membrane potential. 
This exactly reduced model allows for the first time to understand the 
effects of the microscopic neural dynamics at the macroscopic level without making use of deliberate assumptions on the firing statistics.

In this paper, we employ the exactly reduced model introduced in \cite{devalle2017firing} 
for QIF neurons with exponential synapses 
to study the emergence of mixed-mode oscillations in inhibitory networks with fast and slow synaptic kinetics. 
In particular, in sub-section III A we will characterize the dynamical behavior of a single population and review the conditions required to observe collective oscillations (COs). Afterwards, in sub-section III B, we will analyze the response of the population to a modulating signal paying special attention to the emergence of phase synchronization. We will then
consider the possible macroscopic dynamics displayed by two 
coupled populations characterized by different synaptic time scales
in a master-slave configuration (sub-section III C)
and in a bidirectional set-up (sub-section III D). In this latter case we will
focus on the conditions for the emergence of $\theta$-$\gamma$ CFCs  
among the COs displayed by the two populations.

\section{Model and Methods}

\subsection{Network Dynamics}

Through this paper we will consider either one or two populations of 
QIF neurons interacting via inhibitory post-synaptic potentials (IPSPs) with an exponential profile. In this framework, the activity of the population $l \in \{A,B\}$ 
is described by the dynamics of the membrane potentials $V^{(l)}_{i}$ of its neurons and of the associated synaptic fields $S^{(l)}_{i}$.
Here we will assume a fully coupled topology for all networks, hence 
each neuron within a certain population is subject to the same synaptic field
$S^{(l)}$, where the neuron index has been dropped. Therefore,
the dynamics of the network can be written as
\begin{eqnarray}
\label{eq:QIF_network_1pop}
\nonumber \tau \dot{V}^{(l)}_{i} &=& (V^{(l)}_{i})^2 + \eta^{(l)}_{i} + J_{l l}\tau S^{(l)} + J_{k l} \tau S^{(k)} + I^{(l)}(t)\\
		  \tau^{(l)}_{d} \dot{S}^{(l)} &=& -S^{(l)} + \frac{1}{N^{(l)}}\sum_{t^{(l)}_{j}} \delta (t-t^{(l)}_{j}) \\
\nonumber    i &=& 1,...,N^{(l)} \quad l, k \in \{A,B\} \qquad ;
\end{eqnarray}
where $\tau= 10$ ms is the membrane time constant which is assumed 
equal for both populations, $\eta^{(l)}_{i}$ is the excitability of the $i^{th}$ neuron 
of population $l$, $J_{lk}$ is the strength of the inhibitory synaptic coupling of 
population $l$ acting on population $k$ and $I^{(l)}(t)$ is a time dependent external current applied on population $l$. 
The synaptic field $S^{(l)}(t)$ is the linear super-position of the all exponential IPSPs $s(t)={\rm e}^{-t/\tau^{(l)}_d}$ emitted within the $l$ population in the past.
Due to the quadratic term in the membrane potential evolution, which allows the variable to reach infinity in a finite time, the emission of the $j^{th}$ spike in the network occurs at time $t^{(l)}_{j}$ whenever $V^{(l)}_{i}(t^{(l)_-}_{j}) \to +\infty$, 
while the reset mechanism is modeled by setting $V^{(l)}_i(t^{(l)_+}_{j}) \to -\infty$, immediately after the spike emission.\\

For reasons that will be clear in the next paragraph we will also assume that the neuron excitability values $\eta^{(l)}_{i}$ are randomly distributed according to a Lorentzian probability density function  (PDF)
\begin{equation}
\label{eq:lorentzian_eta}
g_l(\eta) = \frac{1}{\pi}\frac{\Delta^{(l)}}{(\eta-\bar{\eta}^{(l)})^2 + (\Delta^{(l)})^2} \;,
\end{equation}
where $\bar{\eta}^{(l)}$ is the median and $\Delta^{(l)}$ is the half-width half-maximum (HWHM) of the PDF. For simplicity, we set $\bar{\eta}=1$ throughout the paper, unless otherwise stated.

In order to characterize the macroscopic dynamics we will employ the following indicators:
$$
r^{(l)}(t)=\frac{1}{N^{(l)} \Delta t}\sum_{t^{(l)}_{j}} \delta (t-t^{(l)}_{j}), \quad v^{(l)}(t) = \frac{1}{N^{(l)}}\sum^{N^{(l)}}_{j} V^{(l)}_j(t),
$$
which represent the average population activity of each network and the average membrane potential, respectively.
In particular the average population activity of the $l-$network $r^{(l)}(t)$ is given by the number of
spikes emitted in a time unit $\Delta t$, divided by the total number of neurons. Furthermore,
the emergence of COs in the dynamical evolution, correspondng to 
periodic motions of $r^{(l)}(t)$ and $v^{(l)}(t)$,  will be
characterized in terms of their frequencies $\nu^{(l)}$.

\begin{figure*}
\includegraphics[width=0.95\linewidth]{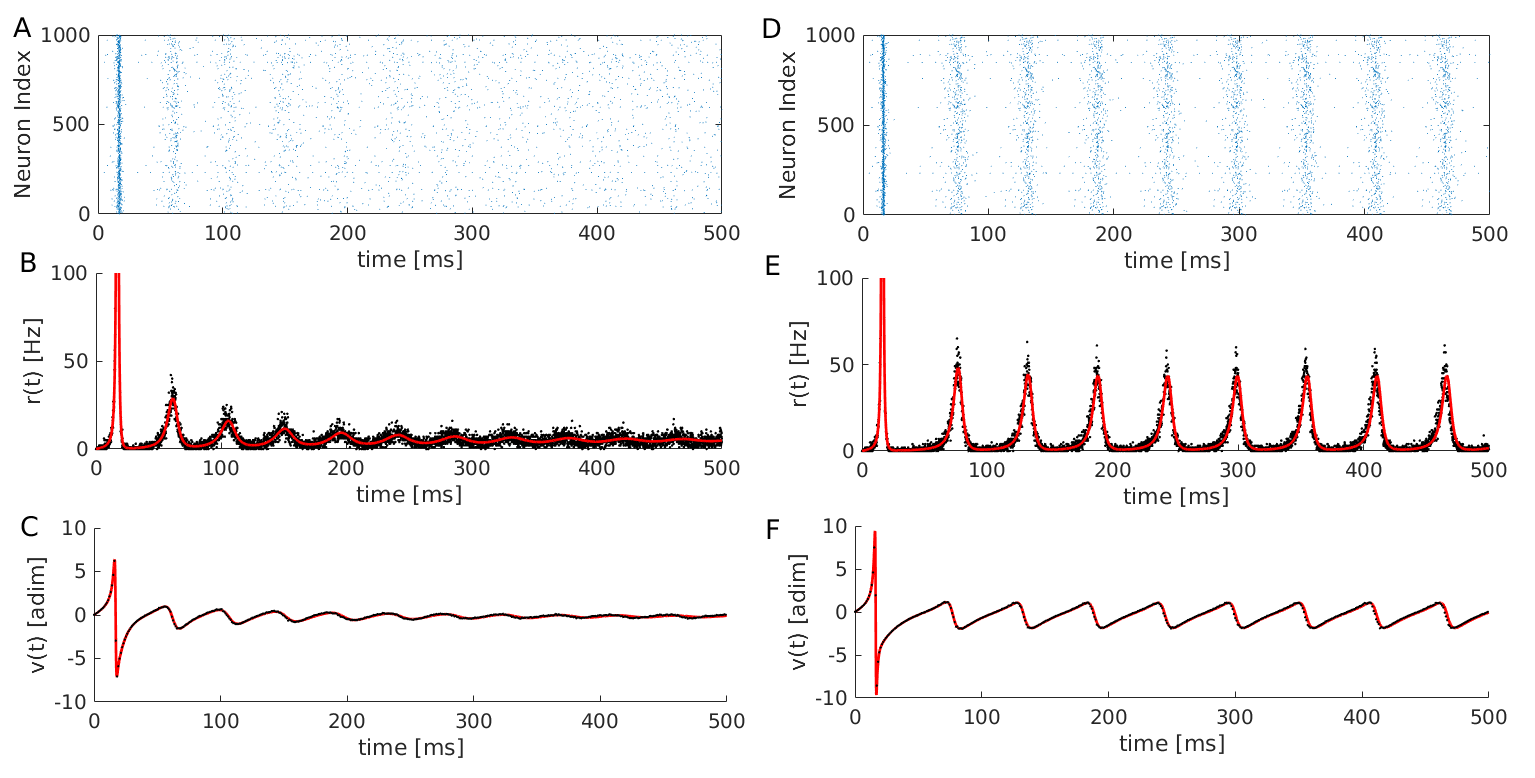}
\caption{\text{\bf Oscillations in a self inhibited QIF population:} 
Left (right) column refers to damped (self-sustained) COs.
From top to bottom: raster plots (A)-(D); istantaneous firing rate $r(t)$ (B)-(E);
average membrane potential $v(t)$(C)-(F). The network simulations are reported as black
dots, while the MF dynamics \eqref{eq:macroscopic_montbrio} is shown as a red line.
For the network simulation $N = 10000$ QIF units are simulated. Left (right) panels were
obtained with $\tau_d = 3$ ms ($\tau_d=8$ ms). Other parameters $\bar{\eta}=1$, $\Delta = 0.05$, 
$J=-20$ and $\tau = 10$ ms. \label{fig:Self_sustained}}
\end{figure*}

\subsection{Mean-Field Evolution:} 

In the limit of infinitely many neurons $N^{(l)} \to \infty$ we 
can derive, for each population $l$, the evolution of the PDF $\rho^{(l)} (V|\eta,t)$ 
describing the probability distribution of finding a neuron with potential $V$ and excitability 
$\eta$ at a time $t$ via the continuity equation:
\begin{eqnarray}
\label{eq:continuity}
\frac{\partial \rho^{(l)}(V|\eta,t)}{\partial t} & = & - \frac{\partial \mathcal{F}^{(l)}(V|\eta,t)}{\partial V^{(l)}}  \\
\nonumber \mathcal{F}^{(l)}(V|\eta,t)&=& \rho^{(l)} \left((V^{(l)})^2 + \eta^{(l)} + \mathcal{I}^{(l)}\right)  \;,
\end{eqnarray}
where $\mathcal{F}^{(l)}(V|\eta,t)$ is the probability flux while $\mathcal{I}^{(l)}= J_{ll}\tau S^{(l)} + J_{kl}\tau S^{(k)} + I^{(l)}(t)$ 
(for $l,k \in \{A,B\}$) is the contribution of all synaptic currents plus the external input.
The continuity equation \eqref{eq:continuity} is completed with the boundary conditions describing the firing and reset mechanisms 
of the network model \eqref{eq:QIF_network_1pop}, which reads as
\begin{eqnarray}
\label{eq:flux_boundary}
\lim_{V \to -\infty} \mathcal{F}^{(l)}(V|\eta,t) = \lim_{V\to \infty} \mathcal{F}^{(l)}(V|\eta,t).
\end{eqnarray}
Following the theoretical framework developed in \cite{montbrio2015macroscopic}, one can apply the 
Ott-Antosen ansatz \cite{ott2008low} to obtain an exact macroscopic description of the infinite dimensional 
2-population system \eqref{eq:QIF_network_1pop} in terms of collective variables. In order to obtain
exact analytic results it is crucial to assume that the distribution of the neuronal excitabilities is a Lorentzian PDF, however the overall picture is not particularly influenced by considering others PDFs, like Gaussian and Erd\"os-Renyi ones \cite{montbrio2015macroscopic}. In the case of the QIF
model the macroscopic variables for each $l-$population are the instantaneous firing rate $r^{(l)}$, the average membrane potential $v^{(l)}$ and the mean synaptic activity $s^{(l)}$, which evolve according to:
\begin{eqnarray}
\label{eq:macroscopic_montbrio}
\dot{r}^{(l)} &=& \frac{\Delta^{(l)}}{\tau^2 \pi} + \frac{2r^{(l)}v^{(l)}}{\tau}\\
\nonumber \dot{v}^{(l)} &=& 
\frac{(v^{(l)})^2 + \bar{\eta}^{(l)} + I^{(l)}(t)}{\tau} +J_{ll} s^{(l)} + J_{kl} s^{(k)}-\tau(\pi r^{(l)})^2 \\
\nonumber 	\dot{s}^{(l)} &=& \frac{1}{\tau^{(l)}_{d}}[-s^{(l)}+r^{(l)}],
\end{eqnarray}
for $l,k \in \{A,B\}$.\\

\subsection{Dynamical Indicators}

We make use of two dynamical indicators to 
characterize the evolution of the MF model \eqref{eq:macroscopic_montbrio}:
a Poincar\'{e} section and the corresponding Lyapunov Spectrum (LS). 

The considered Poincar\'{e} section is defined as 
the manifold $\dot{r}^{(A)}({\bar t}) = 0$ with $\dot{r}^{(A)}({\bar t}^-) > 0$, which amounts 
to identify the local maxima $r_{max}$ of the time trace $r^{(A)}(t)$. 
On the other hand to compute the LS one should consider
the time evolution of the tangent vector $\bm{\delta} = \{\delta r^{(A)}, \delta v^{(A)}, \delta s^{(A)},\delta r^{(B)}, \delta v^{(B)}, \delta s^{(B)} \}$
resulting from the linearization of the original system Eq. \eqref{eq:macroscopic_montbrio}, namely
\begin{eqnarray}
\label{eq:tangent_space}
\delta \dot{r}^{(l)} &=& \frac{2 (r^{(l)}\delta v^{(l)} + v^{(A)} \delta r^{(A)})}{\tau}  \\
\nonumber \delta \dot{v}^{(l)} &=& 
\frac{2v^{(l)}\delta v^{(l)}}{\tau}  + J_{ll} \delta s^{(l)} + J_{kl} \delta s^{(k)} - 2\pi^2 \tau r^{(l)}\delta r^{(l)} \\
\nonumber \delta \dot{s}^{(l)} &=& 
\frac{\delta s^{(l)} + \delta r^{(l)}}{\tau^{(l)}_{d}} \qquad .
\end{eqnarray}
The LS is thus composed by 6 Lyapunov Exponents (LEs) $\{\lambda_i\}$ which quantify the average growth rates of infinitesimal perturbations along 
the different orthogonal manifolds estimated as 
\begin{equation}
\lambda_i = \lim_{t\to\infty} \frac{1}{t}\log \frac{|\bm{\delta}(t)|}{|\bm{\delta}_{0}|} \; ,
\end{equation} 
by employing the well known technique described in Benettin et al. \cite{BenettinLyapunov1980} to maintain the tangent vectors ortho-normal during
the evolution.
From the knowledge of the LS one can obtain an estimation
of the fractal dimension of the corresponding invariant set
in terms of the so-called Kaplan-Yorke dimension $D_{KY}$ \citep{kaplan1979},
defined implicitly as follows:
\begin{equation}
\sum_{i=1}^{D_{KY}} \lambda_i \equiv 0 \; ,
\end{equation}

\subsection{Locking Characterization} 

To investigate the capability of two interacting populations to lock their dynamics in a biologically relevant manner, we measure the degree of phase synchronization
in the exactly reduced system \eqref{eq:macroscopic_montbrio}. 
Therefore, we extract the phase of the population activity,
by performing the Hilbert transform $\mathcal{H}[\cdot]$ of the firing rate for each population $l \in \{A,B\}$, thus obtaining 
the imaginary part of the analytic signal: namely, $\phi^{(l)}(t) := r^{(l)}(t) + j\mathcal{H}[r^{(l)}(t)]$. The evolution of the phase in time is 
then obtained as $\Phi^{(l)}(t) = arg[\phi^{(l)}(t)]$. A generalized phase difference of the $n:m$ phase-locked mode can be defined as:
\begin{equation}
\Delta \Phi_{nm} (t) = n \Phi^{(A)} - m \Phi^{(B)} \quad ,
\end{equation}
and the degree of synchronization in the phase locked regime can be quantified in terms of the Kuramoto order parameter for the phase difference, namely:
\begin{equation}
\label{eq:order_parameter}
\rho_{nm} = |\langle e^{j\Delta \Phi_{nm} (t)}\rangle | \quad ,
\end{equation}
where $\langle \cdot \rangle$ denotes a time average, and $| \cdot |$ is the norm of the complex number \cite{kuramotobook,belluscio2012cross}.

\begin{figure*}
\includegraphics[width=0.95\linewidth]{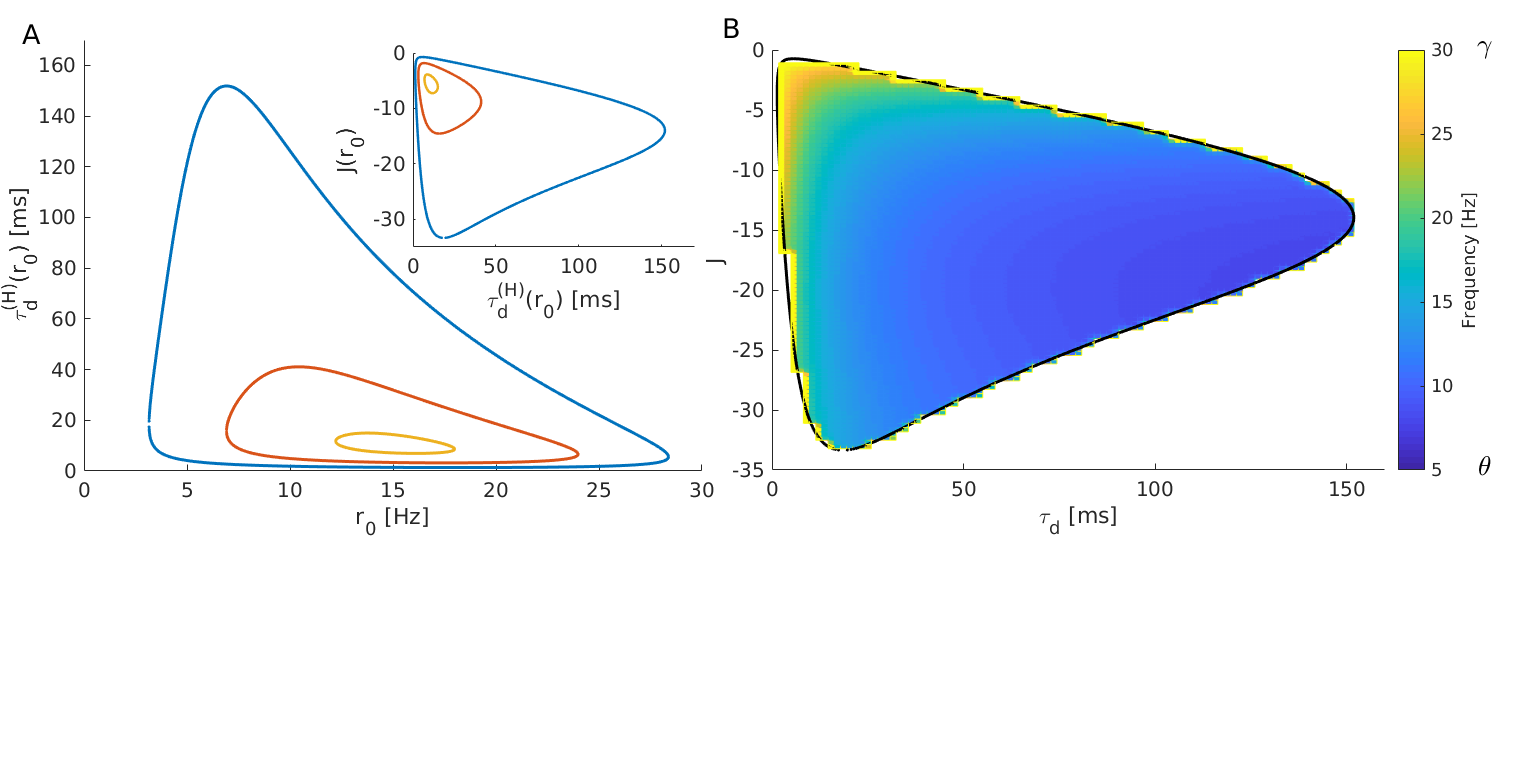}
\vspace{-30mm}
\caption{\textbf{Oscillation Stability Region}: A) Hopf bifurcation boundaries in the $(r_0,\tau_{d})$ plane obtained from \eqref{eq:Hopf_tau} for three different values of the heterogeneity: namely $\Delta = 0.05$ (blue), $\Delta = 0.1$ (red), 
$\Delta = 0.14$ (yellow). Inset: Hopf bifurcation boundaries in the $(\tau_{d},J)$ space for the same  $\Delta$-values. 
B) Heat map of the frequencies of oscillation of the instantaneous firing rate in the $(\tau_d,J)$ plane, for $\Delta = 0.05$. 
The black curve coincides with the Hopf boundary denoted in blue in the inset of panel A). 
To quantify the frequencies of COs, a transient time $t_{t}=1$ s is discarded and then the number of peaks in $r(t)$ are counted and divided by the simulation time $t_s = 2$ s. For this figure, $\tau = 10$ ms and $\bar\eta = 1$.
\label{fig:Hopf_boundaries}}
\end{figure*}

\section{Results}

\subsection{Self sustained oscillations in one population}
\label{sec:onePopulation}

Firstly, by following \citep{devalle2017firing} we analyse the case of a single population with self-inhibition in absence of any external drive. 
Without lack of generality we take in account just population $A$: this amounts to have a set of only three equations in \eqref{eq:macroscopic_montbrio} with $J_{BA}=I^{(A)}(t)=0$. For simplicity in the notation we finally drop the indices denoting the populations.

In the fully coupled QIF network oscillations can be observed only
for IPSPs of finite duration, namely exponential in the
present case. In particular, COs appear when
the equilibrium point of the macroscopic system $({r}_0,{v}_0,{s}_0)$  
undergoes a Hopf bifurcation. 
Simulations of the QIF network model and the corresponding MF dynamics
are compared in Fig. \ref{fig:Self_sustained} revealing a very good agreement
both in the asynchronous and in the oscillatory state.
In particular, for the parameters considered in Fig. \ref{fig:Self_sustained}
the super-critical Hopf bifurcation takes place at $\tau_{d}^{(H)} = 4.95$ms
and panels Fig. \ref{fig:Self_sustained} (A-C) refer to a stable focus for the MF at
$\tau_d < \tau_d^{(H)}$  ,
while panels Fig. \ref{fig:Self_sustained} (D-E) to a stable limit cycle
for $\tau_d > \tau_d^{(H)}$.

Due to the simplicity of the reduced model it is also possible to parametrize the Hopf boundaries 
where the asynchronous state loses stability as a function of the marginally stable solution $({r}_0,{v}_0,{s}_0)$. 
In particular taking $\tau_d$ and $J$ as bifurcation parameters, one can verify that the boundaries of the Hopf 
bifurcation curves are defined by:
\begin{widetext}
\begin{eqnarray}
\label{eq:Hopf_tau}
\tau_{d}^{(H)} & = & \frac{9 \tau  v_0^2 \bar\eta  \tau -\pi ^2 r_0^2 \tau ^3\pm\tau \sqrt{ \left(\bar\eta -\pi ^2 r_0^2 \tau ^2\right)^2+2 v_0^2 \left(9 \bar\eta -41 \pi ^2 r_0^2 \tau ^2\right)+17 v_0^4}}{16 \left(\pi ^2 r_0^2 \tau ^2 v_0+v_0^3\right)}\\
J^{(H)} &=& \frac{\bar\eta + v_0^2}{r_0 \tau }-\pi ^2 r_0 \tau
\end{eqnarray}
\end{widetext}

The equilibrium values are related by the equalities $v_0=-\Delta/(2 \tau\pi r_0)$ and $s_0=r_0$, with $r_0$  acting as a free parameter (see the Appendix for more details).

The phase diagrams showing the existence of self sustained oscillations in the 
$\{r_0,\tau_{d}\}$ plane are displayed in Fig. \ref{fig:Hopf_boundaries} (A), for three values of $\Delta$: 
the region inside the closed curves corresponds to the oscillating regime. 
Upon decreasing the dispersion of the excitability ($\Delta$), the region of oscillatory 
behavior increases. This result highlights that some degree of homogeneity in the neural population
is required in order to sustain a collective activity. In particular, for dispersions larger than a critical value $\Delta_c$ it is impossible for the system to sustain COs (for the parameter employed in the figure $\Delta_c \approx 0.1453$).

The inset in Fig. \ref{fig:Hopf_boundaries} (A) displays the same 
boundaries in the $\{\tau_{d},J\}$ plane. From this figure one can observe that, upon increasing (decreasing) 
$\Delta$, the range of inhibitory strength and of synaptic times required to sustain oscillations decreases (increase).
Thus indicating that more heterogeneous is the system the more the parameters 
$J$ and $\tau_d$ should be finely tuned in order to have COs. 
It's worth mentioning that for instantaneous synapses (corresponding to  $\tau_d \to 0$) no oscillations can emerge autonomously in fully coupled systems 
with homogeneous synaptic coupling as shown in \cite{devalle2017firing} and evident from  Fig. \ref{fig:Hopf_boundaries}.
However, COs can be observed in sparse balanced networks also for instantaneous synapses and in absence of any delay in the signal transmission \cite{divolo}.

In order to understand the role played by the different parameters in modifying the frequency of the COs, we have estimated this frequency for $\Delta=0.5$ in the 
$(\tau_d,J)$-plane. The results are shown as a 
heat-map in Fig. \ref{fig:Hopf_boundaries} (B). 
It turns out that the frequency tends to decrease for increasing values of $\tau_d$ while 
it is almost independent on the value of $J$. 

It is also worth noticing the fundamental role played by the self-inhibition in sustaining 
the autonomously generated oscillations, as it becomes clear from Eq. \eqref{eq:Hopf_tau}, 
since the Hopf bifurcations exist only for negative values of $J$.

From these results we can conclude that a single population of QIF neurons can self sustain 
oscillations with a wide range of frequencies $\nu \simeq 5-30$ Hz thanks to a 
finite synaptic time and to the self-inhibitory action of
the neurons within the population.

\subsection{One population under external forcing}

\begin{figure*}
\includegraphics[width=0.95\linewidth]{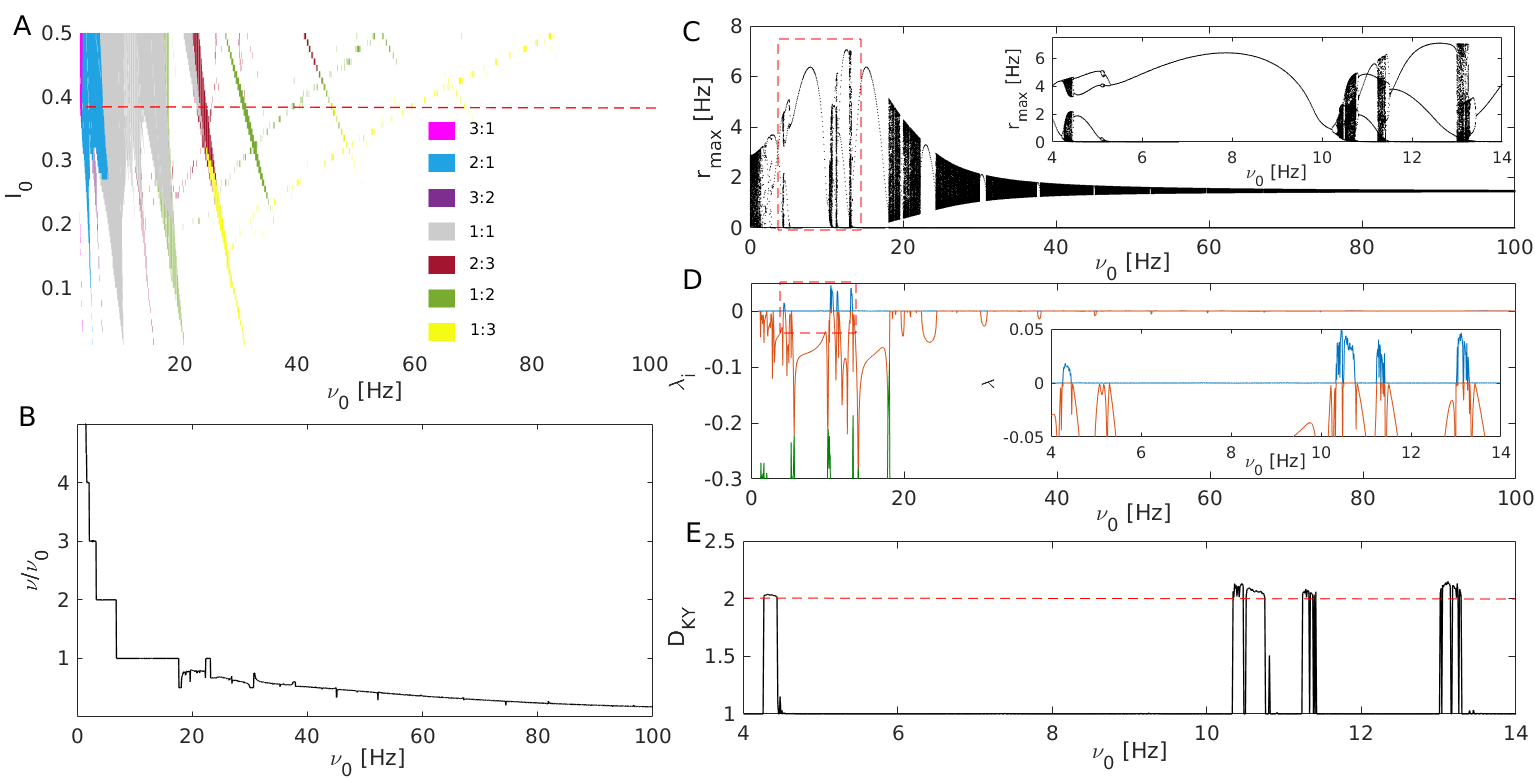}
\caption{\textbf{Single population driven by a harmonic signal}: A) Phase locking diagram of a single inhibitory population driven by a purely inhibitory harmonic external drive \eqref{harm} in the $(I_0,\nu_0)$-plane. The Arnold's tongues 
corresponding to the different regions of phase locking are plotted in colors according to the color code reported in the panel.
The dashed line indicates the value $I_0=0.4$ analysed in the subsequent panels. B) Ratio of the system's oscillation 
frequency $\nu$ and of the driving frequency $\nu_0$ as a function of $\nu_0$ showing a devil's staircase structure. Maxima of $r$ (C) and
the corresponding Lyapunov exponents (D) as a function of $\nu_0$. 
In particular, the first, second and third LEs are reported
as blue, red and green lines, respectively. 
The dashed rectangles indicate the zoomed regions in C) and D) shown in the corresponding insets. 
E) Kaplan-Yorke fractal dimension $D_{KY}$ versus $\nu_0$, the dashed red line
denotes a value of two. The employed parameters are $J = -10$, $\bar{\eta}$, $\Delta = 0.01$, $\tau = 10$ ms, $\tau_d = 80$ ms. For the
estimation of $r_{max}$ and of the Lyapunov exponents a transient time of $t_t = 1$ s was discarded and then the maximum values were stored over a time interval $t_s=3$ s, while for the LS the tangent space was followed for a period $t_s=20$ s. 
\label{fig:Onepop_driven}}
\end{figure*}

Another relevant scenario in the framework of CFC \citep{hyafil2015neural},
is the case where an oscillatory drive $I(t)$ is applied to a neural population
exhibiting COs. The forcing term can represent an input generated from an another neural population
or an external stimulus. Therefore, we examine the behavior of the MF model 
driven by the following harmonic signal 
\begin{equation}
I(t)= -I_0 (1+\sin(2\pi\nu_0 t))
\label{harm}
\end{equation}
characterized by a driving frequency $\nu_0$ and an amplitude $I_0$. 
Notice that we have chosen a strictly negative harmonic signal, to asses the effect 
on the population dynamics of a driving signal originating from a distinct inhibitory population.

The results of this analysis are illustrated in Fig. \ref{fig:Onepop_driven}. First, we study the 
phase locking of the population dynamics, characterized by an oscillatory frequency $\nu$, to the modulatory input, for different forcing frequencies 
$\nu_0$ and amplitudes $I_0$ (see Fig. \ref{fig:Onepop_driven} (A)). 
For small amplitudes, the external modulation is only able to lock the dynamics into a given $n:m$ mode (measured in terms of the indicator \eqref{eq:order_parameter})
for a limited range of forcing frequencies $\nu_0$, while the ratio $n:m$  decreases for increasing $\nu_0$. Furthermore, the range 
where phase locking is observable increases with the amplitude $I_0$, thus giving rise to the Arnold tongues shown in Fig. \ref{fig:Onepop_driven} (A).

\begin{figure}
\includegraphics[width=0.95\linewidth]{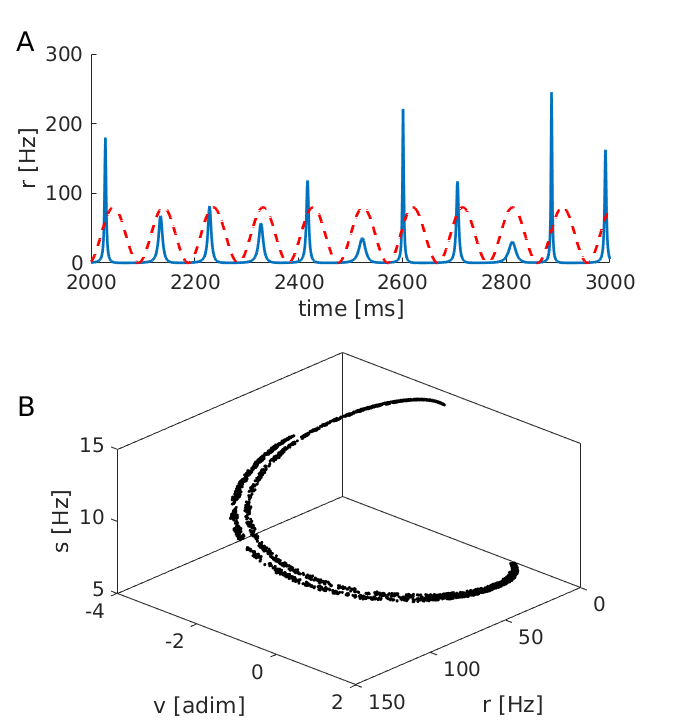}
\caption{\textbf{Chaotic attractor for a single driven population}: A) Time trace of the average firing rate $r(t)$ (blue solid line) and profile of the forcing term $I(t)$ (red dashed line) are shown for a chaotic regime.The chaotic dynamics is clearly locked to the frequency of the external drive.  B)
Stroboscopic attractor: the values of $(r,v,s)$ are recorded at 
regular time intervals corresponding to integer multiples of the forcing period $\nu_0^{-1}$. Parameters for this figure as in Fig. \ref{fig:Onepop_driven} with $\nu_0 = 10.4$ Hz.
\label{fig:Onepop_driven_chaos}}
\end{figure}

To better understand how the locking emerges, we consider the ratio $\nu/\nu_0$
between the CO frequency and the forcing one for a large interval of $\nu_0$-values
and for  a fixed amplitude value $I_0=0.4$, denoted as a red dashed line in Fig. \ref{fig:Onepop_driven} (A). The results are reported in Fig. \ref{fig:Onepop_driven} (B), the ratio $\nu/\nu_0$ 
reveals a structure similar to a devil's staircase, presenting plateaus (corresponding to the locked modes) intermingled with regions where the ratio has not always a monotonic behaviour.
For the same parameters, we also report the maxima $r_{max}$ of the instantaneous firing
frequency of the forced system and the corresponding Lyapunov Spectrum (LS)
in Figs. \ref{fig:Onepop_driven} (C) and (D), respectively. From these 
two indicators one can infer that most of the phase-locked regions correspond to regular periodic motion, as revealed by the single 
value of $r_{max}$  and by a single zero LE observables in a large portion of the devil's staircase plateaus. On the other hand, 
for $\nu_0 > 15$ Hz the regions where the $r(t)$ presents peaks of different heights are in correspondence with the non-flat regions of the devil's staircase.
In particular these regions are associated to quasi-periodic motions, as confirmed by the existence of two zero LEs in the LS. A zoom in the region $\nu_0 = [4,\, 14]$ Hz, corresponding to  $1:1$ locking, is reported in the insets of Figs. \ref{fig:Onepop_driven} (C) and (D). These enlargements show the emergence of 
chaotic windows, where $r_{max}$ assumes values over continuous intervals and
the maximal LE is positive. Furthermore, in these chaotic windows $D_{KY}$ is
slightly larger than two, as shown in \ref{fig:Onepop_driven} (E), indicating
that the chaotic attractor is low dimensional. This is confirmed by
the stroboscopic attractor reported in Fig. \ref{fig:Onepop_driven_chaos} (B),
obtained by reporting the macroscopic variables at regular time intervals equal to integer multiples of the forcing period $\nu_0^{-1}$. 

Indeed the points of the attractor 
cover a set with a dimension slightly larger than one, since one
degree of freedom is lost due to the stroboscopic observation.
Interestingly, the chaotic motion appears despite the $1:1$ locking, this means that the time trace of $r(t)$ presents always a single oscillation within a cycle of the external forcing but characterized
by different amplitudes \cite{pikovsky1997phase}, as shown in Fig. \ref{fig:Onepop_driven_chaos} (A).

\subsection{Two populations in a master-slave configuration}

Despite the fact that at a macroscopic level the network dynamics of a single
population with exponential synapses is exactly described in the
limit $N \to \infty$ by three degrees of freedom \eqref{eq:macroscopic_montbrio}
our and previous analysis \cite{devalle2017firing} have not reported evidences of
chaotic motions for a single inhibitory population.
The situation is different for an excitatory population, as briefly discussed in \cite{bi2019}, or in presence of an external forcing  as shown in the previous sub-section.

In this sub-section we want to analyze the dynamical regimes emerging when a fast oscillating population (indicated as A) is driven by a a slowly oscillating population (denoted by B) in a master-slave configuration corresponding to 
$J_{BA} \ne 0$ and $J_{AB} = 0$. Particular attention will be devoted
to chaotic regimes. The different possible scenarios can be captured by considering 
only two sets of parameters, denoted as $\mathcal{C}_1$ and $\mathcal{C}_2$
and essentially characterized by different ratios of the synaptic time scales of the fast and slow family,
namely:
\begin{widetext}
\begin{eqnarray}
\nonumber \mathcal{C}_1 &:=& \{\tau^{(A)}_{d}=10 \enskip ms, J_{AA} = -10, \tau^{(B)}_{d}=50 \enskip ms, J_{BB}=-16 \} \\
\nonumber \mathcal{C}_2 &:=& \{\tau^{(A)}_{d}=2.5 \enskip ms, J_{AA} = -10,\tau^{(B)}_{d}=80 \enskip ms,  J_{BB}=-20\}.
\end{eqnarray}
\end{widetext}
The coupling between the two population $J_{BA}$ and the network heterogeneity $ \Delta = \Delta^{(A)} = \Delta^{(B)}$
will be employed as control parameters, while we will assume for simplicity $\bar{\eta}^{(A)} = \bar{\eta}^{(B)} = 1$. 

\begin{figure}
\includegraphics[width=0.95\linewidth]{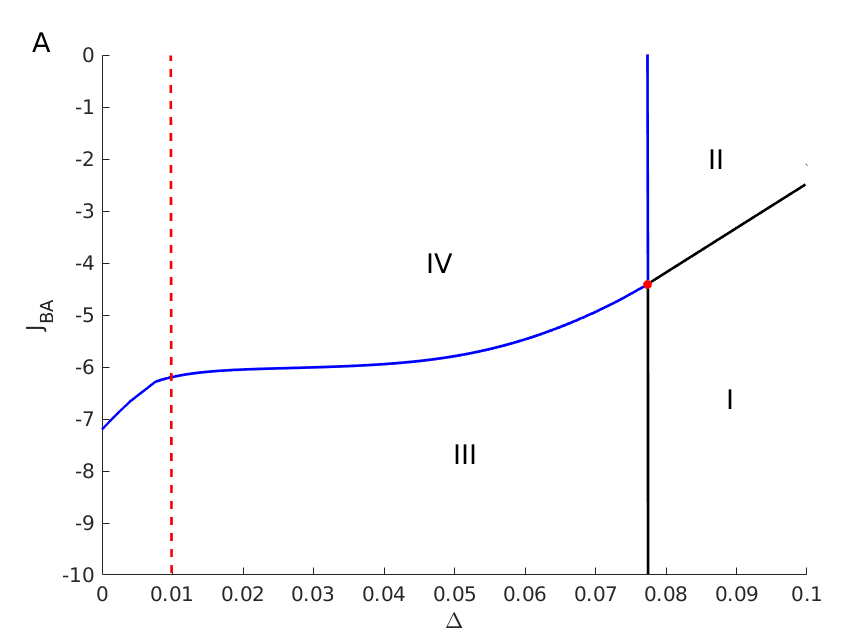}
\includegraphics[width=0.95\linewidth]{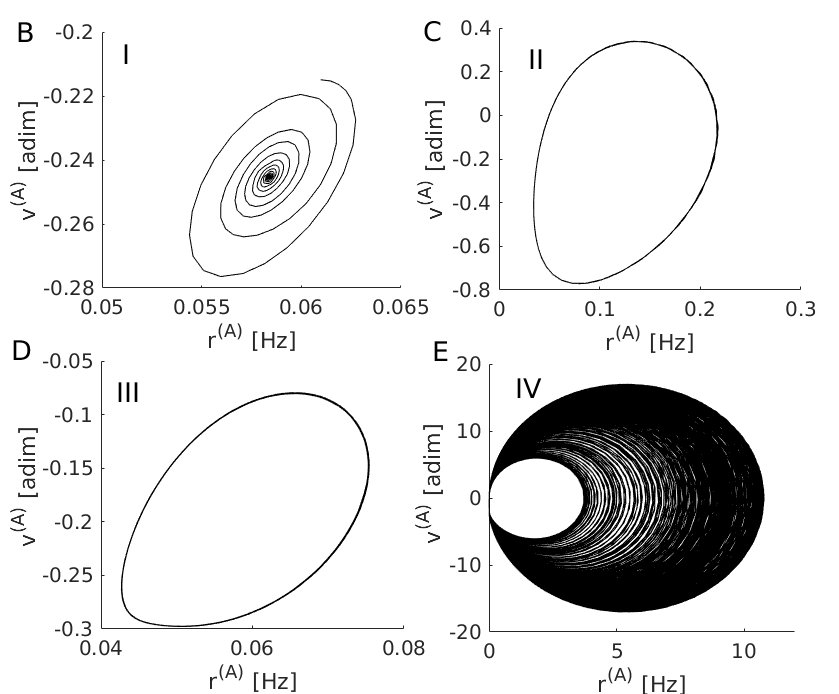}
\caption{\textbf{Bifurcation diagram for the parameter set $\mathcal{C}_1$}. 
A) Bifurcation diagram in the plane $(\Delta , J_{BA})$ for the master-slave 
coupled system. The codimension 2 bifurcation point (red circle) divides the plane in 4 different regions (I-IV) corresponding to different dynamical regimes.
The Hopf bifurcations (black lines) separate foci from limit cycles,
while the Torus bifurcations (blue lines) denote the emergence of Tori $T^2$
from limit cycles. A sample trajectory for each one of the four regions are 
shown in panels (B-E). The red dashed line in (A) indicates the value 
of $\Delta$ considered in Fig. \ref{fig:C1_param_results}. 
\label{fig:C1_param_2D_diagram}}
\end{figure}

Let us first consider the set of parameters $\mathcal{C}_1$, in this case the analysis of the possible bifurcations arising in the $\{\Delta,J_{BA}\}$ plane reveals the existence of a codimension two bifurcation point at 
$(\Delta^{(H)},J_{BA}^{(H)}) \approx (0.078, -4.75)$ which organizes
the plane in four different regions. In these regions, labelled I-IV, the prevalent dynamics corresponds to stable foci in I, stable limit cycles in II and III, 
and to  stable Tori $T^2$ in IV (see Fig. \ref{fig:C1_param_2D_diagram} (A) ). 
For each region we report in  Fig. \ref{fig:C1_param_2D_diagram} (B) a corresponding sample trajectory projected in the sub-space $\{r^{(A)}, v^{(A)}\}$ taken in proximity of the codimension two point.

The critical vertical line observable at $\Delta^{(H)} \simeq 0.077$
in Fig. \ref{fig:C1_param_2D_diagram} (A) is a direct consequence of the super-critical Hopf bifurcation already present at the level of single population discussed in sub-section III A. This line for 
$ J < J_{BA}^{(H)}$ is the locus of Hopf bifurcations
(black solid) dividing foci (I) from stable oscillations (III),
while at larger coupling $J_{BA}$ it becomes a secondary Hopf (or Torus) bifurcation line (blue solid) separating periodic (II) from quasi-periodic motions
(IV). Moreover, the region III of stable limit cycles is  divided by the region IV where Tori $T^2$ emerge from an another Torus bifurcation line (blue solid).
Finally regions I and II are separated by a super-critical Hopf line (black solid).

\begin{figure*}
\includegraphics[width=0.95\linewidth]{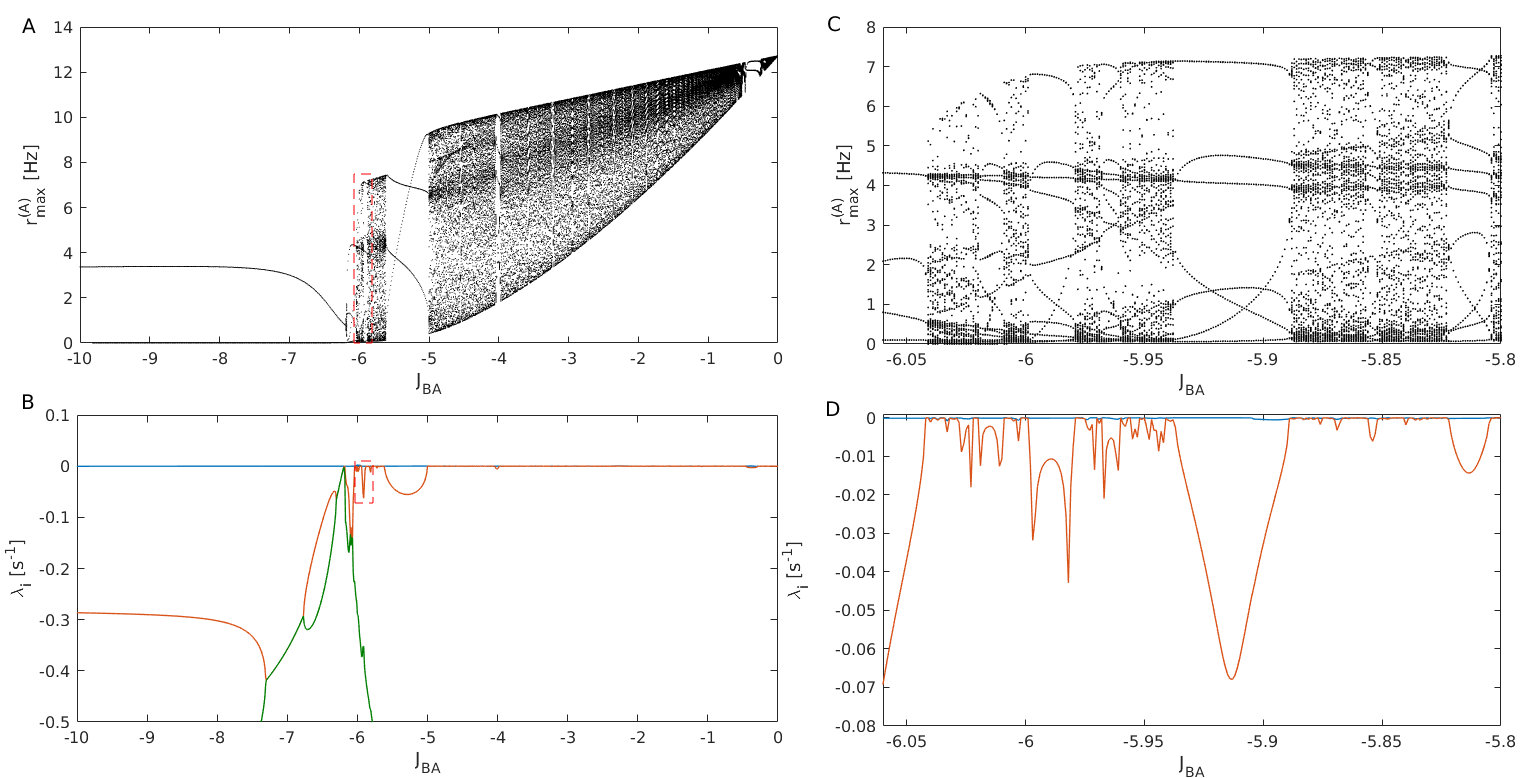}
\caption{\textbf{Characterization of the dynamics for the parameter set $\mathcal{C}_1$}. A) - C) Local maxima $r^{(A)}_{max}$ of the firing rate of the A-population and B) - D) LEs as a function of the coupling strength $J_{AB}$.
In panels (B,D) the blue curve represents the first LE, the orange the second LE and the green one the third LE. The dashed rectangle in panel A (B) indicates the zoomed region presented in panel C (D). For the evaluation of the maxima, a transient time 
of $t_t = 10$ s was discarded and then maximum values were stored during $t_s=15$ s. For the LEs estimation, after 
discarding the transient time $t_t$, the evolution of the tangent space was followed for a period $t=500$ s.
\label{fig:C1_param_results}}
\end{figure*}

We will now focus on the case
$\Delta = 0.01$ (corresponding to the red dashed line in Fig. \ref{fig:C1_param_2D_diagram} (A)), to analyze the different 
regimes observable by varying $J_{BA}$.
To this aim, similarly to what done in the previous sub-sections, we characterize the dynamics of the system in terms of the values of the Poincar\'e map 
$r_{max}^{(A)}$ and of the associated LS.
In particular in Fig. \ref{fig:C1_param_results} (A) are reported the values of
$r_{max}^{(A)}$ in the range of cross-inhibition $J_{BA}=[-10,\,0]$.
At very negative values of the cross-coupling we observe a single 
value for $r_{max}^{(A)}$, which corresponds to periodic COs.
This is confirmed by the values of the LS reported in \ref{fig:C1_param_results} (B): the LEs are all negative except for the first one that is zero. 
At $J_{BA} \approx -6.05$ a broad band appears for the distribution of $r_{max}^{(A)}$ indicating that the time trace now displays maxima of different heights. 
This is due to the Torus bifurcation leading from a periodic to
a quasi-periodic motion. The emergence of quasi-periodic motions is confirmed 
by the fact that in the corresponding intervals the first two LEs
are zero (Fig. \ref{fig:C1_param_results} (D)).
For larger values of the cross-coupling (namely, $J_{BA} \approx [-5.8:-5]$)
a period three window is clearly observable. Beyond this interval
one observes quasi-periodic motions for almost all the negative values
of the cross-coupling, apart narrow parameter intervals were locking of the two
frequencies of the COs occur, as expected beyond a Torus bifurcation \cite{kuznetsov2013}.
 
\begin{figure*}
\includegraphics[width=0.95\linewidth]{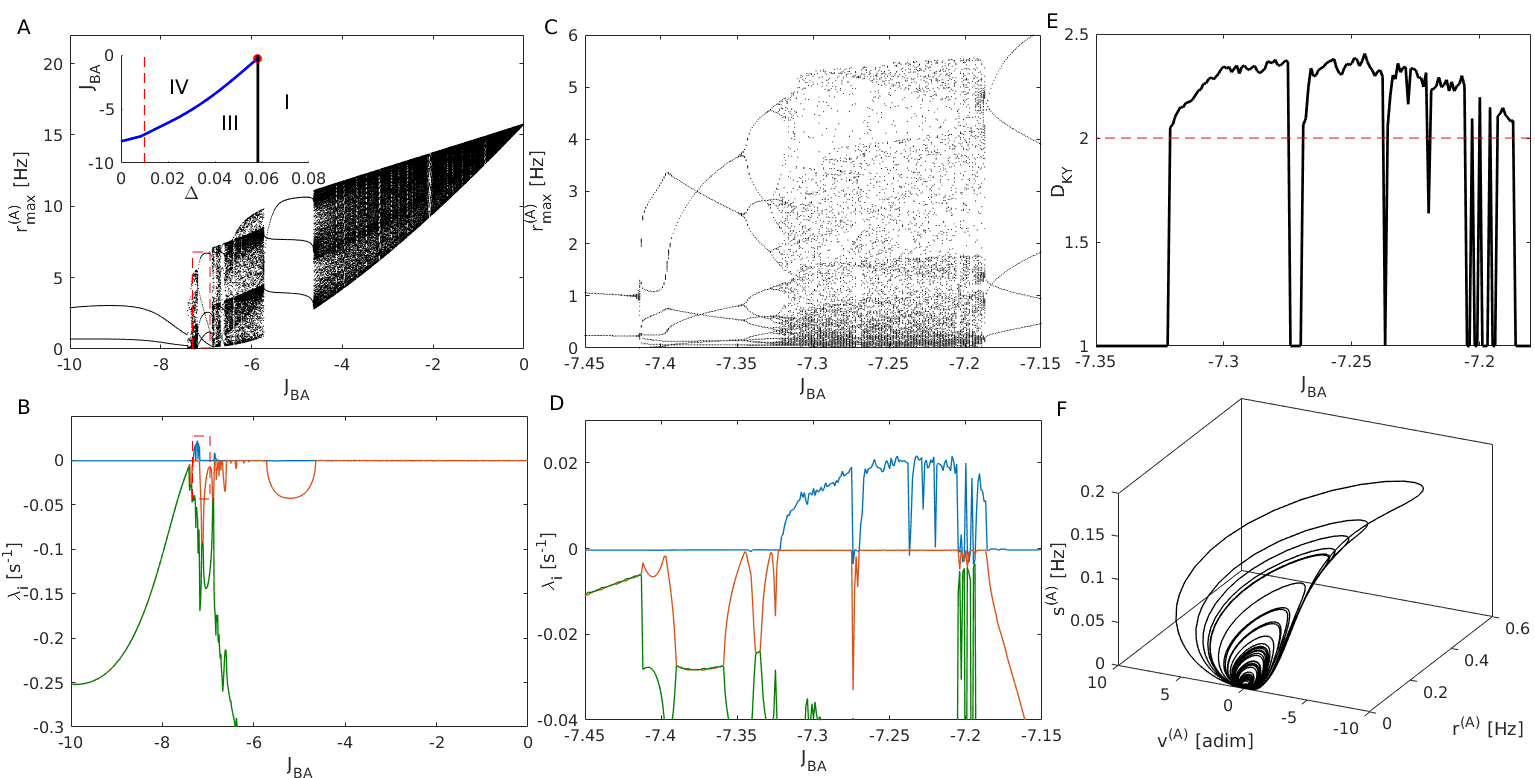}
\caption{\textbf{Characterization of the dynamics for the parameter set $\mathcal{C}_2$}. A-C) Values of $r^{(A)}_{max}$ as a function of $J_{BA}$.
 In the inset it is shown the bifurcation diagram in the $(\Delta,J_{BA})$ plane. This reveals the same structure of the diagram in Fig. \ref{fig:C1_param_2D_diagram} (A), here region (II) is not shown since it corresponds to the region of excitatory cross-coupling. B-D) First three 
LEs as a function of cross-inhibitory coupling $J_{BA}$. 
The red dashed rectangle in panel A (B) denotes the zoomed region presented 
in panel C (D). E) Kaplan-Yorke dimension $D_{KY}$ in the 
parameter interval where chaos is present. F) Chaotic attractor in the $(r^{(A)},v^{(A)},s^{(A)})$ space 
at $J_{BA}=-7.25$. For the evaluation of $r^{(A)}_{max}$, a transient time of 
$t_t = 10$ s was discarded and then maximal values were stored during $t_s=15$ s. 
For the LEs estimation, after the same transient time as before, the evolution of the tangent space was followed during $t=50$ s.
\label{fig:C2_param_results}
}
\end{figure*}

We then proceed to study the parameter set $\mathcal{C}_2$. First, we show that the bidimensional bifurcation
diagram in the plane $\{\Delta,J_{BA}\}$ presents a similar structure to that reported
in Fig. \ref{fig:C1_param_2D_diagram} (A). Indeed also in the present case a codimension two point
located at $\{\Delta^{(H)},J_{BA}^{(H)}\} \approx \{0.06,0\}$ divides the phase space in 4 regions analogous to those observed for the  parameter set $\mathcal{C}_1$ and separated by the same kind of bifurcations (see the inset  of Fig. \ref{fig:C2_param_results} (A)).
As in the previous case, we select the value $\Delta = 0.01$ for analyzing the distribution of maxima of the firing rate $r^{(A)}$ and the associated Lyapunov spectra , see Fig. \ref{fig:C2_param_results}.
For highly negative values of the cross-inhibition we observe a periodic behaviour of the firing rate $r^{(A)}$. At the intersection with the torus bifurcation (occurring at $J_{BA} = -7.48$) quasi-periodicity emerges similarly to what observed for the  parameter set $\mathcal{C}_1$.
However, at larger values of the cross-inhibition (namely $J_{BA} \in [-7.3445, \, -7.3217]$) we observe a period-doubling cascade leading to chaos for $J_{BA} \simeq -7.32$. The systems stays chaotic in the interval
$J_{BA} \in [-7.32, \, -7.18]$ apart for the occurrence of periodic windows.
This is confirmed by the fact that the maximal LE becomes positive
in the corresponding interval, as shown in Fig. \ref{fig:C2_param_results} (D).
An example of chaotic attractor is reported in Fig. \ref{fig:C2_param_results} (F)
with a fractal dimension slightly larger than two, as confirmed
also by the estimation of $D_{KY}$ whose values are
displayed in Fig. \ref{fig:C2_param_results} (E). As in the case of the single
forced population the macroscopic chaotic attractor is low dimensional.
 
For $ J_{BA} > -7.18$ we have periodic and quasi-periodic activity, but no more chaos, in particular for $J_{BA} > -4.2$ we essentially observe mostly quasi-periodic motions up to $J_{BA} = 0$.
As mentioned before, the main difference between the parameter sets $\mathcal{C}_1$ and $\mathcal{C}_2$ is the ratio of the time scales associated to the synaptic filtering. For the parameter set $\mathcal{C}_1$ the time scale ratio $\tau_{d}^{(A)}/\tau_{d}^{(B)}$ is 1:5, while for the set $\mathcal{C}_2$ becomes 1:32. 
We have been able to find chaotic motions only for inhibitory populations
with this large difference in their synaptic time scale. 
However, these values are biologically plausible, indeed
they can correspond to populations of
interneurons generating IPSPs mediated via GABA$_{A,fast}$ and GABA$_{A,slow}$ receptors \cite{1}, which have been identified in
the hippocampus \cite{banks1998} and in the cortex \cite{sceniak2008}.

\subsection{Cross-frequency-coupling in bidirectionally coupled populations}
 
As already mentioned a fundamental example of CFC, is represented by the coupling of the $\theta$ and $\gamma$ rhythms. Gamma oscillations are usually modulated 
by theta oscillations during locomotory actions and rapid eye movement (REM) sleep 
in the hippocampus \cite{lisman2013theta} as well as in the neocortex \cite{sirota2008}.  While gamma oscillations have been shown to
be crucially dependent on inhibitory networks \cite{buzsaki2012mechanisms},
the origin of the $\theta$-modulation is still under debate. It has been
suggested to be due either to an external excitatory drive \cite{buzsaki2002} or to a cross-disinhibition originating from a distinct inhibitory population \cite{white2000networks,hangya2009}.
 
In this sub-section we analyze the possibility that two bidirectionally interacting inhibitory populations could be at the basis of the $\theta-\gamma$ CFC.  
Inspired by previous analysis, we propose the difference in the synaptic 
time kinetics as a possible mechanism to achieve CFC \cite{white2000networks}.
Therefore we set the synaptic time scale of the
fast (slow) population to $\tau_{A,d}=9$ ms 
($\tau_{B,d}=50$ ms), which corresponds approximately to the 
time scales of IPSPs generated via GABA$_{A,fast}$ (GABA$_{A,slow}$) receptors. Regarding the other parameters, internal to each population, these are 
chosen in a such a way that the self-generated oscillations correspond 
roughly to $\theta$ and $\gamma$ rhythms, respectively.

\begin{figure*}
\includegraphics[width=0.95\linewidth]{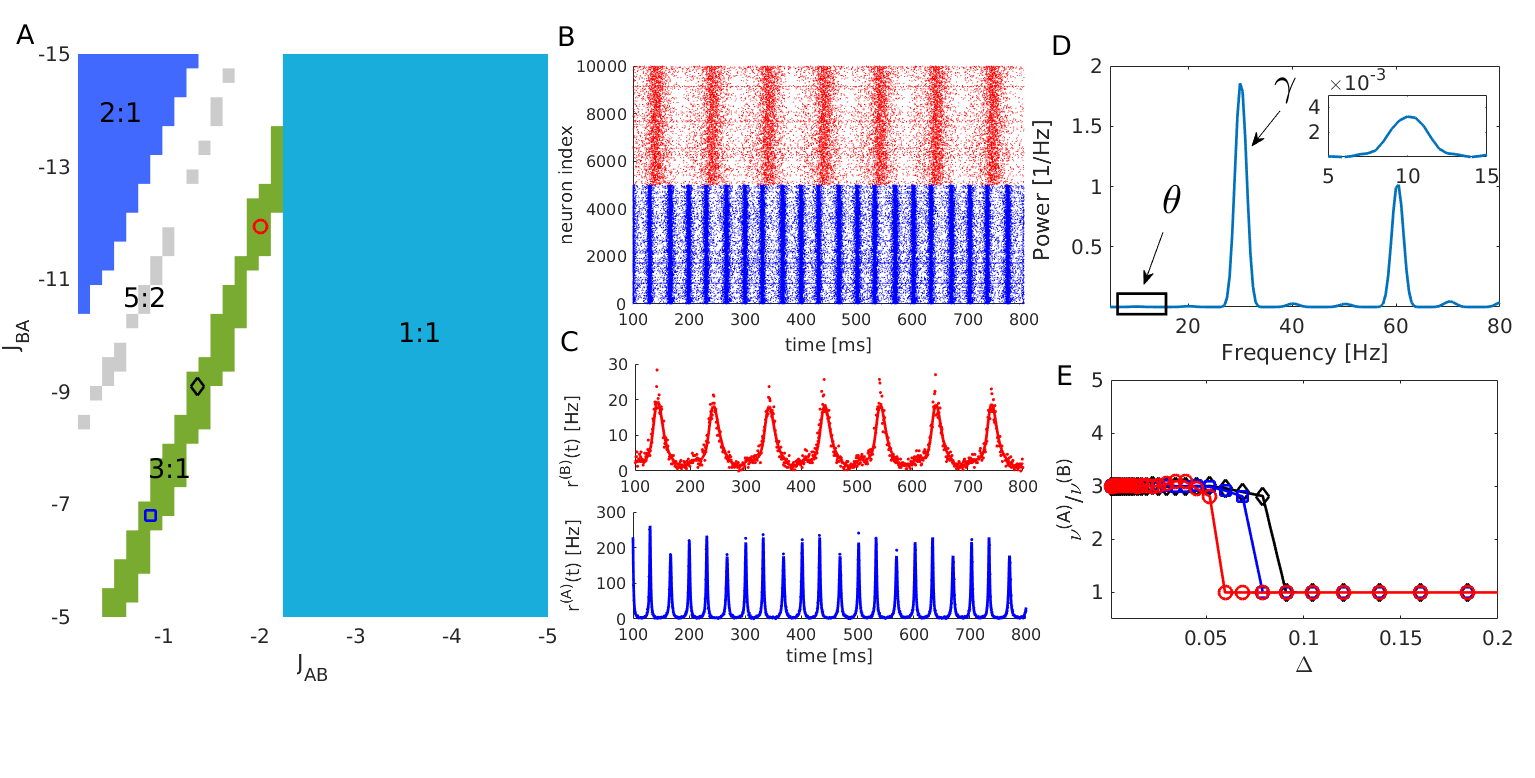}
\vspace{-10mm}
\caption{\textbf{CFC in bidirectionally coupled populations}. A) Heat map of the order parameter locking modes for different values of the cross-coupling. Colored symbols denote three couple of parameters $\{J_{AB},J_{BA}\}$ corresponding to $3:1$ locking examined in (E) for different disorder values.
In particular the red circle denotes the couple $\{J_{AB},J_{BA}\}$
for which one obtains the maximal value of $\rho_{31}$ and these parameter
values are employed for the simulations reported in (B,C). 
B) Raster plot of the network model Eq. \eqref{eq:QIF_network_1pop} 
showing the fast and slow population in blue and red colors respectively. 
C) Instantaneous firing rates of the two populations
obtained from the evolution of the MF dynamics \eqref{eq:macroscopic_montbrio} 
(same color code as in panel (B)). 
D) Power spectrum of the time trace $r^{(A)}(t)$ shown as a blue line
in panel (C), in the inset an enlargement of the spectrum corresponding to the
$\theta$-band is shown. E) Ratio of the fast and slow frequencies of the COs
$\nu^{(A)}/\nu^{(B)}$ showing the extent 
of the $3:1$ locking interval for the three values of $\{J_{AB},J_{BA}\}$ 
denoted by the symbols of the same color in (A) versus the 
disorder $\Delta := \Delta_A = \Delta_B$. Parameters for fast population are
$\tau_{d}^{(A)}=9\, ms, \bar{\eta}_A = 2,\, J_{AA}= -2$ and for the slow one $\tau_{d}^{(B)}=50 \enskip ms, \bar{\eta}_B=1.5 ,\, J_{BB}=-18$.
Common parameters $\Delta = 0.05$ (for panels A to D) and $\tau = 10$. 
The time traces used for the phase locking analysis were taken over
a period of $t=10$ s after discarding an initial transient time $t_t=10$ s. 
For the network simulations in B), $N_A = N_B = 10000$ neurons, in the figure only half of them are depicted. 
The spectrum in (D) was obtained by averaging 50 power spectra 
each calculated over a time trace of duration $t=16.384$ s
with $2^{15}$ equispaced samples, after a transient time of $t_t=10$ s. 
\label{fig:CFC_noMod}}
\end{figure*}

Firstly, we consider the case in which no external modulation is present, i.e $I^{(l)}(t)=0$.  Depending on the value of the cross-coupling parameters
$J_{AB}$ and $J_{BA}$ different type of $m:n$ P-P coupling can be achieved. In particular, as shown in Fig. \ref{fig:CFC_noMod} (A), we observe
$1:1$ and $2:1$ phase synchronization
in large regions of the  $(J_{AB};J_{BA})$ plane,  while $3:1$ and $5:2$ locking emerge only along restricted stripes of the plane.
In particular, we focus on the values of cross-inhibition $J_{lk}$
for which it is possible to achieve a $3:1$ phase synchronization, corresponding 
to a $\theta-\gamma$ coupling. As evident from the green area in Fig. \ref{fig:CFC_noMod} (A), this specific P-P coupling occurs only for low values of
$J_{AB}$, namely $J_{AB} \in [0.5, 2.5]$. 

Among the parameter values corresponding to $3:1$ locking
we choose for further analysis the ones for which 
the order parameter $\rho_{31}$ is
maximal (denoted as a red circle in Fig. \ref{fig:CFC_noMod} (A)).
In particular, we performed simulation of the
network \eqref{eq:QIF_network_1pop}  as well as of the 
corresponding MF model \eqref{eq:macroscopic_montbrio}.
The raster plot in Fig. \ref{fig:CFC_noMod} (B) confirms
that the two populations display COs locked in a $3:1$ fashion. 
During the burst emitted from the slow population the fast one displays 
irregular asynchronous activity followed by three rapid bursts 
(each lasting around 10 ms), before the next CO of the slow population.
The fact that the slow population emits bursts of longer
duration is confirmed by the analysis of the instantaneous firing rates reported 
in Fig. \ref{fig:CFC_noMod} (D).
Indeed, $r^{(A)}$ has oscillation of amplitude
much larger than $r^{(B)}$ indicating that more neurons are 
recruited for a burst of population (A) with respect to population (B).
This difference in the oscillations amplitude can also explain why the $3:1$ locked mode is observable for $J_{AB} \ll J_{BA}$, indeed for larger
$J_{AB}$ the activity of the slow population would be silenced.
It is worth to notice in Fig. \ref{fig:CFC_noMod} (D)
the good agreement between the firing rates obtained from
the network simulations (dots) and from the evolution
of the MF model (line).

An analysis of the power spectrum of $r^{(A)}$ (shown in
Fig. \ref{fig:CFC_noMod} (D)) reveals that the amount of 
power in the $\theta$ band is quite small
(see the peak around 10 Hz in the inset) with respect to
the power in the $\gamma$ band. This indicates that a CFC
among the two bands is indeed present, but the interaction 
is limited.

Furthermore, varying the amplitude of the heterogeneity, as measured by $\Delta = \Delta^{(A)} = \Delta^{(B)}$, we can verify the capability of the network to sustain the $3:1$ locked mode even in presence of disorder in the neural excitabilities. It can be seen that the system loses 
the ability to sustain such a locked state already for $\Delta > 0.1$, 
indicating that CFC can occur only for a limited 
amount of disorder in the distribution of the neuronal excitabilities
in agreement with the results reported in \cite{white2000networks} 
(see Fig. \ref{fig:CFC_noMod} (E). For large disorder the only possible
locked state is that corresponding to $1:1$ phase synchronization.

Sofar we have analyzed the possibility that $\theta$ and $\gamma$ rhythms were locally generated in inhibitory populations with different synaptic scales.
However, the results of several optogenetic experiments 
performed for different area of the hippocampus and of the enthorinal cortex
suggest that a $\theta$ frequency drive is sufficient to induce {\it in vitro}
$\theta$-$\gamma$ CFCs \cite{pastoll2013,akam2012, butler2016,butler2018}. However, 
the interpretation of these experiments disagrees on the
origin of the locally generated $\gamma$ oscillations.
Two mechanism have been suggested: namely, either inhibitory \cite{pastoll2013,akam2012} or excitatory-inhibitory feedback loops \cite{butler2016,butler2018}.
Therefore, to clarify if recurrently coupled inhibitory populations, with different
synaptic time scales, under a $\theta$-drive can display $\theta$-$\gamma$ CFC
we drive the slow population via an external current $I^{(B)} = I_0^{(B)} \sin(2\pi \nu_{\theta} t)$ with $\nu_{\theta} = 10$ Hz, while the rest of the parameters remains unchanged. 

\begin{figure}
\includegraphics[width=0.95\linewidth]{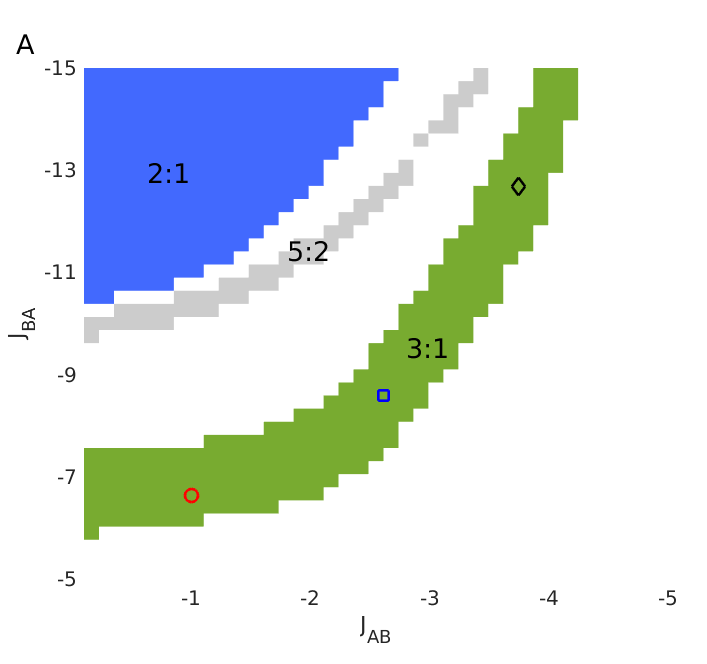}
\includegraphics[width=0.95\linewidth]{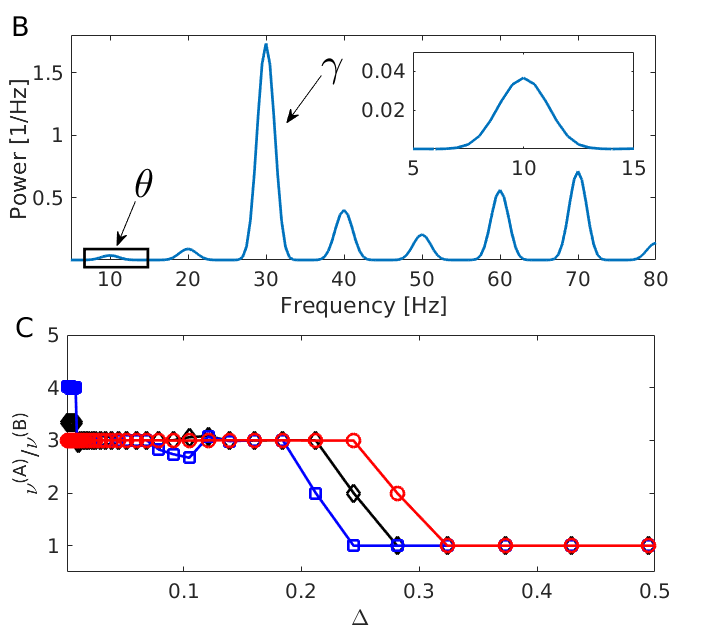}
\caption{\textbf{CFC in bidirectionally coupled populations with an external $\theta$ modulation}. 
A) Locking modes for different values of the cross-coupling. 
Red circle denotes the pair $\{J_{AB},J_{BA}\}$ giving rise to the maximum value of $\rho_{31}$ which is used for simulations in B-C). Black diamond and blue square correspond also to 3:1 modes with smaller order parameter value. B) Power spectrum of the time trace $r^{(A)}(t)$ for the the case denoted by the red circle 
in panel (A), the inset displays an enlargement corresponding to the $\theta$-band. C) Ratio of the fast and slow population frequencies $\nu^{(A)}/\nu^{(B)}$ showing the extend of the 3:1 mode for the 
three values of $\{J_{AB},J_{BA}\}$ depicted by the symbols in A) at varying values of disorder $\Delta := \Delta^{(A)} = \Delta^{(B)}$. 
For the $\theta$-forcing current we set $I_0^{(B)}=0.5$ and $\nu_\theta = 10$ Hz,
all the other parameters as in Fig. \ref{fig:CFC_noMod}.
\label{fig:CFC_ThetaMod}
}
\end{figure}

As before, we look for the range of cross-inhibitions in which a $3:1$ phase-locked mode emerges: results are plotted in Fig. \ref{fig:CFC_ThetaMod} (A). 
We observe that the region where $\theta-\gamma$ CFC can be observed definitely
enlarge in presence of an external $\theta$-modulation.
Furthermore, as observable from the power spectrum 
reported in Fig. \ref{fig:CFC_ThetaMod} (B)
the power in the $\theta$ band is noticeably increased 
as a consequence of the external modulation with respect to the non-modulated case.
Moreover, the $\theta-\gamma$ CFC is now observable over a wider 
range of disorder on the excitabilities, indeed the $3:1$ locked state
survives up to  $\Delta \approx 0.2-0.3$, as shown in Fig. \ref{fig:CFC_ThetaMod} (C). These evidences are similar to what reported in \cite{white2000networks}, where adding a slow modulation to the $\theta$ generating population
was sufficient to render more robust the observed CFC to the presence of disorder in the excitability distribution.
 
\begin{figure}
\includegraphics[width=1.\linewidth]{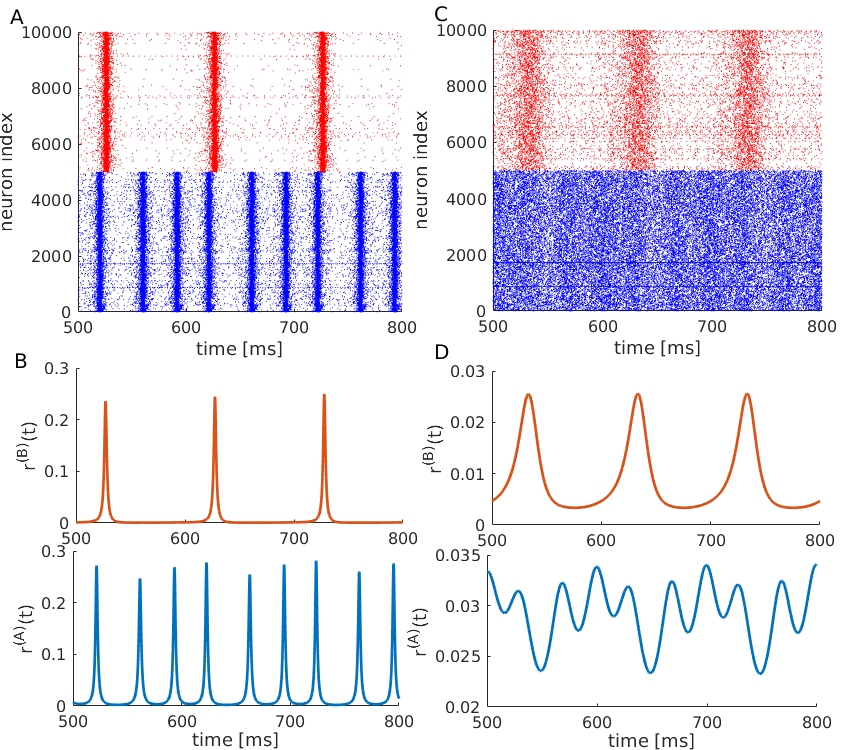}
\caption{\textbf{P-P and P-A couplings in presence of a $\theta$ forcing}. 
A) Raster plot of the network model Eq. \eqref{eq:QIF_network_1pop} showing the fast and slow population in 
blue and red colors respectively for the same system analyzed in Fig. \ref{fig:CFC_ThetaMod} with the optimal
pair $\{J_{AB},J_{BA}\}=\{-1, -6.63 \}$ and $\Delta = 0.05$. B) Time traces of $r(t)$ for the fast (blue) and slow (red) populations
for the case depicted in the raster plot in A). C-D) same as in A-B) for the same parameters except for $\Delta = 0.2$.
Parameters as in Fig. \ref{fig:CFC_ThetaMod}. 
\label{fig:nested_freq}}
\end{figure}

Finally, if we look at the network activity we observe two different
scenarios corresponding to the $3:1$ locked mode: (1) a P-P locking at
low disorder and (2) a P-A locking (or $\theta$-nested $\gamma$ 
oscillations) at larger disorder. The first scenario is characterized
by the fast population displaying clear COs slightly modulated in their 
amplitudes by the activity of the slow population, but tightly locked in phase
with the slow ones (see Fig. \ref{fig:nested_freq} (A,B)).
The second scenario presents a firing activity
of the fast population strongly modulated in its amplitude
by the slow population as observable in Fig. \ref{fig:nested_freq} (A,B).
In this latter case the neurons in population (A) fire almost
asynchronously with a really low firing rate, however the coupling
with by the activity of the slow population (B) is reflected in a clear
modulation of the firing rate $r^{(A)}$, analogously to what reported
for $\theta$-nested $\gamma$ oscillations induced by 
optogenetic stimulation \cite{pastoll2013,butler2016}.

\section{Concluding remarks}

Self sustained oscillatory behavior has been widely studied in the neuroscientific community, as several coding mechanisms 
rely on the rhythms generated by inter-neuronal populations \cite{wang2010neurophysiological}.
In particular, the emergence of COs in inhibitory networks has been usually related to the presence of an additional timescale, 
beyond the one associated with the membrane potential evolution, which can be either the transmission delay \cite{BrunelHakim1999,Brunel2000Sparse} or a finite synaptic time \cite{van1994inhibition,luke2013,devalle2017firing,coombes2019}.
The only example of COs emerging for instantaneous synapses and in absence of delay has been reported in \cite{divolo} for
a sparse network of QIF neurons in a balanced setup. In this case,
for values of the in-degree sufficiently large
the COs emerge due to endogenous fluctuations, which persist in 
the thermodynamic limit.

In this paper we have considered an heterogeneous inhibitory population with exponentially decaying synapses, which can be described at a macroscopic level by an exact reduced model of three variables: namely, the firing rate, the average membrane potential and the mean synaptic activity.
As shown in \cite{devalle2017firing,coombes2019}, the presence of the synaptic dynamics is at the origin of the COs, emerging via a super-critical Hopf bifurcation. In particular, we have shown that the period of the COs is controlled 
by the synaptic time scale and that, for increasing heterogeneity, the observation of COs require finer and finer tuning of the model parameters. 
Moreover, we have characterized in detail the effect of an inhibitory periodic current on a single self-oscillating population. The external forcing
leads to the appearance of locking phenomena characterized by Arnold tongues and devil's staircase. We have also identified low dimensional chaotic windows, where the instantaneous firing rate of the
forced population display oscillations of irregular amplitudes, but tightly locked to the external signal oscillations \cite{pikovsky1997phase}.

Furthermore, we considered two inhibitory populations connected in a master-slave configuration,
i.e. unidirectionally coupled, where the fast oscillating population is forced by the one with slow
synaptic dynamics. In a single population we observed, for sufficiently large heterogeneity, only focus solutions at the macroscopic level, which can turn in COs by reducing the disorder in the network. In presence of a second population the complexity of the macroscopic
solutions increases for cross-inhibitory coupling not too negative. In particular, by increasing the cross-coupling, the focus becomes a limit cycle via a super-critical Hopf bifurcation and the COs a quasi-periodic motion
via a Torus bifurcation. Depending on the parameter values a period doubling cascade leading to chaotic 
behaviour is observable above the Torus bifurcation. In particular, the macroscopic attractor has a fractal dimension slightly larger than two with a single associated positive Lyapunov exponent despite the fact that the two coupled
neural mass model are described by six degrees of freedom. 
The macroscopic solutions we have found in the master-slave configuration are similar to those identified in \cite{luke2014macroscopic} for $\theta$-neurons populations coupled via pulses of finite width.

Even though the dichotomy between chaos and reliability in brain 
coding is a debated topic \cite{GuillaumeChaosBalanced2013,London2010,JhankeTimme2008PRL,angulo2014}, 
it is of great importance to establish the conditions in which such behavior may appear. In the present case, chaotic behavior is found only when the synaptic time scales of the two inhibitory populations are quite different. However,  
these values are consistent with synaptic times observable in interneuron populations with fast and slow GABA$_A$ receptors, as shown in \cite{banks1998,white2000networks,sceniak2008}.
 
We finally explored the possibility that two interacting inhibitory networks could give rise to 
specific cross-frequency-coupling mechanisms \cite{hyafil2015neural}. In particular, for its
relevance in neuroscience we limited the analysis to $\theta$-$\gamma$ CFC
emerging as a consequence of the interaction between fast and slow GABA$_A$ kinetics, as reported 
in \cite{white2000networks} for populations of Hodgkin-Huxley neurons. 
In the original set-up was possible to observe $\theta$-$\gamma$ CFC coupling in a narrow region of parameters
and for a limited range of heterogeneity. The addition of a $\theta$-forcing
on the slow population renders the system more robust to the disorder in the excitability distribution and enlarges the observability region of the $\theta$-$\gamma$ CFC. Furthermore,  we observed two kind of CFC: namely, P-P (P-A) coupling for low (high) heterogeneity.
Both these scenarios have been reported experimentally for $\theta$-$\gamma$
oscillations: namely, P-A coupling has been reported {\it in vitro} 
for optogenetic $\theta$-stimulations of the the hippocampal area CA1 \cite{butler2016} and CA3 \cite{akam2012}, as well as of the medial enthorinal cortex \cite{pastoll2013};  P-P coupling have been observed in the hippocampus in behaving rats \cite{belluscio2012cross,colgin2009}.

Our analysis shows, for the first time to our knowledge, that the $\theta$-$\gamma$ CFC, reported for Hodgkin-Huxley networks in \cite{white2000networks}, can be reproduced also at the level of exact neural mass models for two coupled
inhibitory populations. Our results pave the way
for further studies of other CFC mechanisms present in the brain, 
e.g by employing analytically estimated macroscopic Phase Response Curves \cite{dumont2017macroscopic} to characterize phase synchronization in multiscale networks of QIF neurons.

\acknowledgments

Authors are in debt with Ernest Montbri\'o for various enlightening interactions in the first phase of development of this project, furthermore they
acknowledge fruitful discussions with Federico Devalle, Boris Gutkin, and Alex Roxin. A.T. received financial support by the Excellence Initiative A*MIDEX (Grant No. ANR-11-IDEX-0001-02) (together with D. A.-G.), by the Excellence Initiative I-Site Paris Seine (Grant No ANR-16-IDEX-008), by the Labex MME-DII (Grant No ANR-11-LBX-0023-01) (together with S.O.) and by the ANR Project ERMUNDY (Grant No ANR-18-CE37-0014), all part of the French programme ``Investissements d'Avenir''. D.A-G was also supported by CNRS for a research period at LPTM, UMR 8089, Universit\'e de Cergy-Pontoise, France. A.C was supported by Erasmus+ Traineeship 2016/2017 contract between University of Florence, Department of Mathematics and Computer Science ``Ulisse Dini'' (DIMAI), and Centre de Physique Th\'eorique (CPT) and LabEx Archim\`ede, Marseille, France.

\appendix

\section*{Appendix: Hopf boundaries}

In the case of a single population of inhibitory neurons with no external input,
the fixed point solutions $(v_0,r_0,s_0)$ of the MF model \eqref{eq:macroscopic_montbrio}
are given by the following set of equations
\begin{eqnarray}
\label{eq:special0} v_0 =& -\frac{\Delta}{2 \tau \pi r_0}\\
\label{eq:special}  v_0^2 + \bar\eta - (\pi \tau r_0)^2 + \tau J s_0  =& 0\\
\label{eq:special_2} s_0  =&  r_0 \;.
\end{eqnarray}

In order to study the linear stability of the equilibrium point, we consider 
the corresponding eigenvalue problem, namely
\begin{equation}
\text{det} \left[
\begin{array}{ccc}
 2 v_0/\tau -\Lambda  & 2 r_0/\tau & 0 \\
 -2 \pi^2\tau r_0 & 2 v_0/\tau -\Lambda  & J \\
 1/\tau_d & 0 & -\Lambda - 1/\tau_d \\
\end{array}
\right] = 0
\end{equation}
where $\Lambda$ are the complex eigenvalues which can be found
by solving the following characteristic polynomial
\begin{equation}
p(\Lambda)= \tau_d \tau^2 \Lambda ^3 + A \Lambda ^2 + \Lambda \left(\tau_d B-4 \tau  v_0\right)
+(B-2r_0 J \tau)
\end{equation}
where $A = (\tau^2-4v_0\tau_d\tau)$ and $B=(4 v_0^2  + 4\pi^2r_0^2\tau^2$). 

In order to obtain a parametrization of the Hopf bifurcation curve we 
impose $\Lambda = i \Omega$ with $\Omega \in \Re \setminus \{0\}$ and solve $p(i\Omega)=0$, which can only be satisfied if 
\begin{eqnarray}
\label{eq:Re_omega1}
\text{Re}[p(i\Omega)]=0 \quad \text{and} \quad \text{Im}[p(i\Omega)]=0\;.
\end{eqnarray}

Solving for $\Omega$ Eqs. \eqref{eq:Re_omega1} we end up with:
\begin{eqnarray}
\label{eq:Re_omega2}
\Omega_{\text{Re}}&=&\pm\sqrt{\frac{(B-2r_0 J\tau)}{A}} \\
\label{eq:Im_omega2}
\Omega_{\text{Im}}&=&\pm\sqrt{\frac{(\tau_d B -4 \tau v_0)}{\tau_d \tau^2}} .
\end{eqnarray}

By equating \eqref{eq:Re_omega2} and \eqref{eq:Im_omega2} we can find the values of $J^{(H)}$ where the
Hopf bifurcation occurs, namely
\begin{equation}
\label{eq:JH_app}
J^{(H)} = \frac{2 v_0 \left[\tau ^2 \left(4 \pi ^2 \tau_d ^2 r_0^2+1\right)+4 \tau_d ^2 v_0^2-4 \tau_d  \tau  v_0 \right]}{\tau_d r_0 \tau ^2} \quad .
\end{equation}

Finally, by introducing Eq. \eqref{eq:JH_app} in \eqref{eq:special}, and noticing from Eq. \eqref{eq:special_2} that $s_0=r_0$,
one can derive the values of the  synaptic time scale $\tau_d^{(H)}$ that bounds the oscillating 
region and that is reported in Eq. \eqref{eq:Hopf_tau}. Notice that the dependence on $\Delta$ is implicitly introduced
in the expression of $v_0$, therefore, it possible to obtain also the critical value 
$\Delta^{(H)}$ associated to the Hopf transition by substituting \eqref{eq:special0} in \eqref{eq:special} and solving for $\Delta$.
It should be also stressed that
this approach cannot distinguish between super-critical and sub-critical Hopf bifurcations. As a matter of fact for an inhibitory QIF population with exponential synapses no sub-critical bifurcations have been reported for homogeneous synaptic
couplings \citep{devalle2017firing}, while these emerge whenever a disorder is introduced either in the synaptic couplings or in the link distribution \citep{bi2019}.


\begin{thebibliography}{}
\bibitem[1, ]{1}
GABA (gamma-Aminobutyric acid) is the main inhibitory neurotransmitter in the
  adult mammalian brain, GABA performs its action by binding to GABA$_A$ or
  GABA$_B$ receptors.
\end{thebibliography}

\end{document}